\documentclass[prl,twocolumn,showpacs,preprintnumbers,amsmath,amssymb]{revtex4}

\usepackage{graphicx}% Include figure files
\usepackage{dcolumn} % Align table columns on decimal point
\usepackage{bm}      % bold math

\newlength{\figlength}
\begin{document}

\title{Adaptive sampling by information maximization}

\author{Christian K. Machens}
\email{c.machens@itb.biologie.hu-berlin.de}
\homepage{http://itb.biologie.hu-berlin.de/~machens}
\affiliation{Innovationskolleg Theoretische Biologie,
Invalidenstr. 43, Humboldt-University Berlin, 10115 Berlin, Germany
}

\date{\today}

\begin{abstract}

The investigation of input-output systems often requires a sophisticated choice of test inputs to make best use of limited experimental time. Here we present an iterative algorithm that continuously adjusts an ensemble of test inputs online, subject to the data already acquired about the system under study. The algorithm focuses the input ensemble by maximizing the mutual information between input and output. We apply the algorithm to simulated neurophysiological experiments and show that it serves to extract the ensemble of stimuli that a given neural system ``expects'' as a result of its natural history.
\end{abstract}



\maketitle

Biophysical systems often have many degrees of freedom and thus one needs large numbers of variables and parameters to describe them. Without strong prior knowledge about the intrinsic dynamics of such a system, one is left with inferring its function from data obtained by experiments or observations. Given a system where we control a set of ``input'' variables $x=(x^{(1)},x^{(2)}, \ldots, x^{(n)})$ and measure another set of ``output'' variables $y=(y^{(1)},y^{(2)},\ldots, y^{(m)})$, we can actively manipulate the data acquisition by selecting the most informative test inputs. Yet how should one choose the test inputs to learn most about the input-output relation?

Within the classical Volterra-Wiener system identification methods \cite{Wie:58}, the input space is sampled by drawing inputs from a probability distribution $p(x)$; a common choice is Gaussian ``white noise''. However, not all aspects of the system's input-output relation may be equally important. In neurobiology, for instance, one is especially interested in inputs $x$ about which a given sensory system conveys most information. In the spirit of importance sampling \cite{ItD:89b}, one might therefore focus the data acquisition on those $x$ that contribute most to the information transfer. For a given input distribution, the information provided by a single input can then be quantified as $I(x)=H_y - H_y(x)$ where $H_y$ is the entropy of the output distribution $p(y)$ and $H_y(x)$ is the entropy of the conditional probabilities $p(y|x)$ which characterize the input-output relation \cite{ShW:49,CoT:91}. Hence, the appropriate focusing is achieved by an input distribution $p_{\text{opt}}(x)$ that maximizes the mutual information $I = \langle I(x) \rangle$ where the angular brackets denote averaging over $p_{\text{opt}}(x)$.

Without any information about the system and its input-output relation, the optimal input distribution $p_{\text{opt}}(x)$ is unknown. Any experimental test of the system must therefore start with drawing the test inputs from some predefined distribution $p_{\phi}(x)$ that depends on a set of parameters $\phi = (\phi^{(1)}, \ldots, \phi^{(L)})$. Once data about the system has been acquired, however, one need not adhere to this initial choice of an input distribution. Instead, one should adapt the parameters or even the structure of $p_{\phi}(x)$ to better focus on the important inputs. In this letter, we show how to systematically perform this adaptation. By iterating the adaptation procedure, the acquired data becomes ever more useful and the input distribution approaches the optimum.

%%%%%%%%%%%%%%%%%%%%%%%%%%%%%%%%%%%%%%%%%%%%%%%%%%%%%%%%%%%%%%
{\em Adapting the input distribution}---For mathematical simplicity, we assume that both input and output take discrete values. Say that we have already tested the system with $N$ different inputs $x_i$ each of which was presented $M_i$ times while measuring the outputs $y_{ij}$ with $i=1\ldots N$ and $j=1\ldots M_i$. We define the set of all different output values measured so far by $\{y_k:k=1\ldots K\}$. Our present knowledge about the system is summarized by the conditional probability that an output~$y_k$ was obtained from the input~$x_i$,
\begin{equation}  
q(y_k|x_i) = \frac{1}{M_i} \sum_{j=1}^{M_i} \delta_{y_{ij},y_k} \;.
\label{eq:histogram}
\end{equation}

%Evaluating inputs
The estimated probabilities $q(y_k|x_i)$ allow us to re-evaluate the relative importance of the inputs $x_i$ in terms of their potential contribution to the mutual information. To measure this contribution, we assign a probability or ``weight'' $q(x_i)$ to every input. Initially we assume that all inputs $x_i$ contribute equally and set $q_1(x_i)=1/N$. To find a combination of weights that maximizes the information transfer, we use the Blahut-Arimoto algorithm \cite{Ari:72} and readjust the weights,
\begin{equation}  
q_{n+1}(x_i) = \frac{1}{Z}  q_n(x_i) \exp\Big( \sum_{k=1}^{K}
q(y_k|x_i)  \log  \frac{q(y_k|x_i)} {q_n(y_k)} \Big).
\label{eq:BA} 
\end{equation}
Here $q_n(y_k)=\sum_{i=1}^N q(y_k|x_i) q_n(x_i)$ and $Z$ is a normalization constant so that $\sum_{i=1}^N q_{n+1}(x_i) =1 $. According to Eq.~(\ref{eq:BA}), the weight of an input $x_i$ is decreased if its conditional output distribution $q(y_k|x_i)$ is similar to the total output distribution $q_n(y_k)$. In contrast, the weight of an input $x_i$ is increased if the respective distributions differ. When Eq.~(\ref{eq:BA}) is iterated, the weights converge and reach a global maximum of the mutual information \cite{Ari:72}. In practice, we terminate the process once $ |1 - q_{n+1}(x_i)/q_n(x_i)| < \epsilon$ for all $i$ and some chosen precision $\epsilon$ and set $q_{\text{opt}}(x_i) = q_{n+1}(x_i)$.

%Re-estimating parameters
The weights or probabilities $q_{\text{opt}}(x_i)$ describe the relative frequencies with which the respective inputs $x_i$ should be drawn. Consequently, we need to adapt the parameters $\phi$ so as to find a matching distribution $p_{\phi}(x)$. Here we determine the new parameters~$\phi$ by maximizing the log-likelihood function \cite{Bar:89a},
\begin{equation}
        \log L(x_1, \ldots, x_N | \phi) = \sum_{i=1}^N q_{\text{opt}} (x_i) \log p_{\phi}(x_i)
\label{eq:mle}
\end{equation}
where the probabilities $q_{\text{opt}}(x_i)$ provide the appropriate weights. For some model distributions, e.g.~Gaussians, the maximum can be found analytically. In general, however, one has to evaluate the maximum numerically.

The input distribution given by the new parameter values can be used to draw new test inputs, present them to the system and measure the respective outputs. After a certain amount of data has been acquired, the parameters $\phi$ of the input distribution can be adapted again. The resulting iterative algorithm moves the input distribution towards an optimal ensemble.

%%%%%%%%%%%%%%%%%%%%%%%%%%%%%%%%%%%%%%%%%%%%%%%%%%%%%%%%%%%%%%
{\em Model quality and convergence}---Every maximum of the mutual information with respect to $p(x)$ is a global maximum \cite{CoT:91}. Hence, if the model distribution does not rule out any inputs, i.e., $p_{\phi}(x)>0$ for all $x$ and $\phi$, the estimates of the input-output relation, Eq.~(\ref{eq:histogram}), converge, and therefore $q_{\text{opt}}(x_i) \to p_{\text{opt}}(x_i)$. Accordingly, the mutual information $ I_{\text{D}} = \langle H_y^q - H_y(x) \rangle_q $ achieves the information capacity of the system; here, the index $q$ denotes that the respective quantities and averages are calculated with respect to $q_{\text{opt}}(x_i)$.

The model distribution $p_{\phi}(x)$ converges towards an optimal fit of $p_{\text{opt}}(x)$. To control how well the model distribution captures the structure of the optimal distribution, one can check the mutual information achieved by the model, $I_{\text{M}} = \langle H_y^{\phi} - H_y(x) \rangle_{\phi}$, which is calculated with respect to  $r_{\phi}(x_i) =  p_{\phi}(x_i) / [\sum_{j=1}^N p_{\phi}(x_j)]$. The fraction $\gamma$ of the mutual information captured by the model  is then defined as
\begin{equation}
  \gamma = \frac{I_{\text{M}}}{I_{\text{D}}}
\end{equation}
and provides a measure for the quality of the model. Hence, if $\gamma$ falls significantly below one, the model does no longer capture the structure of the optimal ensemble; in such a case, one might increase the complexity of the model.

In general, the algorithm will not be able to adapt the input ensemble if the presented inputs always result in the same output value. Similarly, there is no possibility to weight the inputs $x_i$ differently if every input elicits a new, different output. However, the latter problem can be solved by discretizing the output side into a smaller number of possible outputs. The input space, on the other hand, can be discretized as fine as needed without impeding the convergence of $p_{\phi}(x)$.

%%%%%%%%%%%%%%%%%%%%%%%%%%%%%%%%%%%%%%%%%%%%%%%%%%%%%%%%%%%%%%
{\em Example}---To illustrate the method, we study a numerical simulation of a Hodgkin-Huxley-type model neuron \cite{WaB:96}. The model neuron transforms an input current~$I$ into a voltage output~$V$. For constant current values $I<0\;\mu\text{A/cm}^2$, the voltage approaches a stable equilibrium. For current values $I>0\;\mu\text{A/cm}^2$, the model undergoes a saddle-node bifurcation and generates periodically occurring action potentials, also called spikes \cite{Izh:00}. Stochastic aspects of neural activity are incorporated by adding Gaussian white noise with a fixed standard deviation $\sigma_{\eta}$ and a cut-off frequency $f_{\eta}$ to the input. 

\begin{figure}[t]

\includegraphics[width=\figlength]{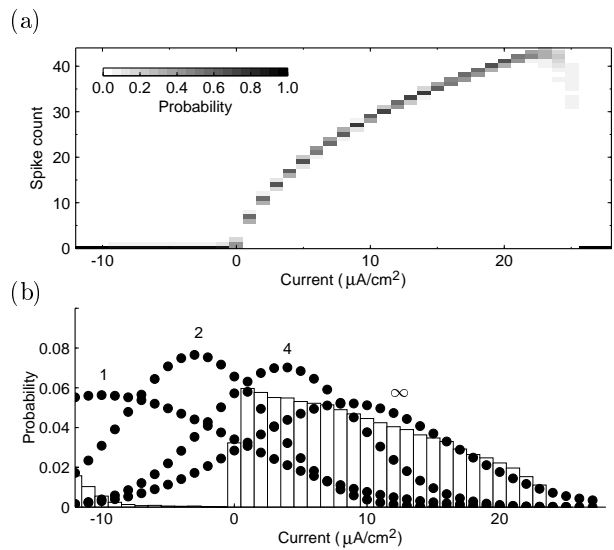}

\caption{\label{fig:onedim}
Approaching the optimal input ensemble of a neuron with static, one-dimensional input and output.
(a)~Plot of the conditional probability distribution $p(C|I)$ with spike count~$C$ and input current~$I$. The uncertainties at $I \approx 22\;\mu\text{A/cm}^2$ are due to a decline in spike size that makes it impossible to detect the spikes in the noisy voltage output. For $I\approx 28\;\mu\text{A/cm}^2$, the model neuron ceases to generate spikes.
(b)~Approaching the optimal input distribution (bars). Shown are the initial distribution (1), the distributions of the iterations (2) and (4), as well as the final distribution ($\infty$).
(Simulation parameters: $n=1$, $m=1$, $L=2$, $A=10$, $B=5$, $\epsilon = 0.1$, $\sigma_{\eta} = 4\;\mu\text{A/cm}^2$, $f_{\eta}=1000$~Hz)
}
\end{figure}

%Parametrisierung
We start with a simple one-dimensional parametrization of input and output. The inputs are 100-ms-long, discretized current steps ($\Delta I=1\;\mu \text{A/cm}^2$), restricted to a physiologically realistic range of $I=-12\ldots 28\;\mu\text{A/cm}^2$. The outputs are given by the number of spikes, $C$, during the corresponding time window. The resulting probabilistic relation of spike count versus current is displayed in Fig.~\ref{fig:onedim}(a).

%Exakte Loesung
For this one-dimensional input-output system, we can compute an exact solution of the information maximization problem. The optimal input distribution $p_{\text{opt}}(I)$ is depicted by the vertical bars in Fig.~\ref{fig:onedim}(b); the shape of $p_{\text{opt}}(I)$ corresponds to the slope of the input-output relation \cite{footnote}. Note that there is a slight increase in the probabilities of inputs far below threshold ($I \leq -10 \;\mu \text{A/cm}^2$). These inputs result almost certainly in a zero spike count output. At the same time, inputs closer to threshold ($I\approx -9\ldots -1 \;\mu \text{A/cm}^2 $) are more likely to produce spikes. As the optimal input distribution favors inputs that are more reliable, the inputs closer to threshold are neglected.

%Gauss-Approximation der Loesung
To study the performance of the iterative algorithm, we model the optimal input distribution by a truncated Gaussian. As initial parameter values, we choose a mean $\phi^{(1)}=-10 \;\mu \text{A/cm}^2 $ and a standard deviation $\phi^{(2)}=10 \;\mu \text{A/cm}^2 $. In each iteration, we draw $A$ current values from the Gaussian, test them $B$ times on the system, and adapt the parameters. For our Gaussian model, the maximum likelihood estimate of the new parameters is given by $\phi^{(1)} = \sum_{i=1}^N  I_i q_{\text{opt}}(I_i)$ and $\phi^{(2)} = \big[ \sum_{i=1}^N ( I_i - \phi^{(1)} )^2 q_{\text{opt}}(I_i) \big]^{1/2}$.

The Gaussian model distributions are displayed in Fig.~\ref{fig:onedim}(b) for the first few iterations. Most of the current values drawn from the initial distribution fall below the spiking threshold of the neuron. Consequently, the algorithm shifts the Gaussian distribution into the spiking regime of the neuron. After about 10~iterations, the mutual information rate saturates at $\approx 40$~bits/sec. Since both the final Gaussian model $p_{\phi}(I)$ and the optimal input distribution $p_{\text{opt}}(I)$ lead to approximately the same information transfer, the landscape of the mutual information with respect to the input distribution is relatively flat around the maximum; it suffices if the input distribution covers the relevant input range ($I\approx 0\ldots 25\;\mu\text{A/cm}^2$). Note, that due to the maximum-likelihood estimation, Eq.~(\ref{eq:mle}), the final Gaussian distribution has the same mean and variance as the optimal distribution.

%%%%%%%%%%%%%%%%%%%%%%%%%%%%%%%%%%%%%%%%%%%%%%%%%%%%%%%%%%%%%%
{\em Multi-dimensional example}---The computational power of the algorithm becomes clearly visible for high-dimensional input spaces. As an example, consider the above model neuron when the input consists of time-varying, statistically stationary currents, discretized in time steps of $\Delta t_1$. Following \cite{SKR+:98}, we slide overlapping windows of length $T=n\Delta t_1$ across the input current trace and use the values within each window as input vector $I_i = (I_i^{(1)}, \ldots, I_i^{(n)})$. For each of these inputs $I_i$, the output $C_{ij}$ is given by the spike times, discretized in time steps of $\Delta t_2 = T/m$, during the corresponding window. Hence, each input consists of $n$ real-valued numbers bounded within the interval $I=-12\ldots 28\;\mu\text{A/cm}^2$, and each output consists of $m$~numbers whose values are either zero (no spike) or one (spike). Note, that we do not explicitly discretize the current values; we instead assume that every input $I_i$ is unique. For simplicity, we use a Gaussian input distribution. As the input is real and stationary, it suffices to use $L=n/2+1$ parameters for describing average and power spectrum of the current trace.

\begin{figure}[t]

\includegraphics[width=\figlength]{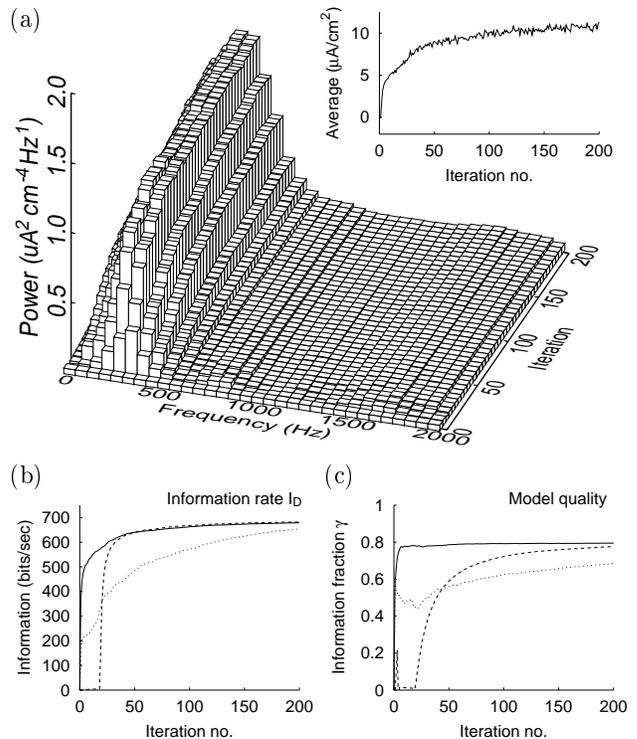}

\caption{\label{fig:multi}
Approaching the optimal input ensemble of a neuron with time-varying input and output.
(a)~Evolution of average and power spectrum. 
(b)~Evolution of information rate and (c) model quality for three different initial conditions.
(Simulation parameters: $n=64$, $m=16$, $L=33$, $A=1000$, $B=20$, $\epsilon = 0.1$, $\sigma_{\eta} = 4\;\mu\text{A/cm}^2$, $f_{\eta}=1000$~Hz, $\Delta t_1 = 0.25$~ms, $\Delta t_2=1$~ms, $T=16$~ms; windows slided by $\Delta t_2$; accordingly, $A\Delta t_2 B \times 100$~iterations $\approx 34$~minutes)
}
\end{figure}

To test the system, we choose an initial distribution with an average $\phi^{(1)} = 0\;\mu\text{A/cm}^2$ and a flat power spectrum with standard deviation $\sigma = \big[\sum_{i=1}^{n/2}\phi^{(i)}\big]^{1/2} = 10\;\mu{A/cm}^2$. For this prior, only 50\% of the input values lie above threshold and the inputs will rarely lead to high firing rates, cf.~Fig.~\ref{fig:onedim}. Consequently, we do not properly explore the full range of the input-output relation; if, for example, we test the system for 30~minutes with input currents drawn from this initial distribution, the information rate $I_{\text{D}}$ does not exceed $\approx 300$~bits/sec.

When using the iterative algorithm to adapt the parameters of the input ensemble, on the other hand, the information rate $I_{\text{D}}$ saturates around $\approx$~670 bits/sec after about 20~minutes. Figure~\ref{fig:multi}(a) shows how the power spectrum is shaped during the iterations. Only input frequencies below 500~Hz are well suited for the information transfer, the cut-off is roughly determined by the maximum firing rate of the model neuron. The overall increase in power leads to input currents that override the additive noise $\eta$ of the model neuron.

%%%%%%%%%%%%%%%%%%%%%%%%%%%%%%%%%%%%%%%%%%%%%%%%%%%%%%%%%%%%%%%%%%%%%%%%%
{\em Initial conditions, convergence, and degeneracies}---When the initial distribution is very narrow (flat power spectrum up to $f_c=1000~Hz$, with $\sigma=1\;\mu\text{A/cm}^2$, $\phi^{(1)}=20\;\mu{A/cm}^2$, Fig.~\ref{fig:multi}(b), dotted line), most of the input currents drive the neuron maximally and thereby very reliably. The strong initial bias leaves the algorithm with little maneuvering space for the parameter re-estimation so that it takes longer to approximate an optimal input distribution.

In the worst case, every input leads to the same output value. With the initial choice $\phi^{(1)}=-6\;\mu\text{A/cm}^2$ and $\sigma = 10\;\mu\text{A/cm}^2$, the input does not elicit any spikes during the first iterations, cf.~Fig.~\ref{fig:multi}(b), dashed line. However, once a spike has appeared, the statistics of the model distribution immediately moves into the direction of the statistics of the inputs~$I_i$ that caused a spike. When the algorithm has tracked the relevant input range, a rapid increase of the information rate follows.

In the examples studied, the mutual information reaches approximately the same value independent of the initial conditions, cf.~Fig.~\ref{fig:multi}(b). Although there is always a clear preference for frequencies below 500~Hz, however, the parameters of the optimal input ensemble do not converge to the same set of values. Consequently, there is no unique combination of parameters that maximizes the mutual information; an observation that generalizes beyond the specific examples chosen. Nonetheless, the final input distributions always capture about $\gamma=80$\% of the mutual information $I_{\text{D}}$, cf.~Fig.~\ref{fig:multi}(c).

In general, there might be ``degenerate'' subsets in stimulus space, i.e., sets of stimuli that lead to the same output value. In these cases, the total probability assigned to such a subset can be distributed in an arbitrary way on the subset and any statistical parameters~$\phi$ that depend on these subsets can assume different values without significant consequences for the information transfer.

%%%%%%%%%%%%%%%%%%%%%%%%%%%%%%%%%%%%%%%%%%%%%%%%%%%%%%%%%%%%%%%%%%%%%%%%%
{\em Neurophysiological interpretation}---Recent studies indicate that sensory neurons convey large amounts of information if the properties of the stimulus ensembles used match those of natural stimuli \cite{RBB:95}. Here we have shown how to extract a stimulus ensemble that conveys the maximum possible information without any prior knowledge. The proposed method could therefore serve to find the ensemble of stimuli that a given neuron naturally ``expects''. Note that in contrast to previous online algorithms such as Alopex or Simplex \cite{HaT:74}, we are not looking for a single optimal stimulus but rather for a complete ensemble of stimuli.

Our results demonstrate also that the optimal stimulus ensemble depends on the chosen criteria about what aspect of the output carries the relevant information. Hence, if the investigated model neuron conveys information in its average firing rate, then it best encodes slow-varying current values in the range $I=0\ldots23\;\mu\text{A/cm}^2$. Synaptic inputs should therefore drive the neuron in the corresponding range. If, on the other hand, a neuron encodes its information in the precise timing of spikes, then the synaptic input should be of a more binary nature and either fully excite or fully inhibit the neuron. Measuring the time courses of a neuron's membrane potential thus allows conclusions about the used neural code under optimal conditions.

\begin{acknowledgments}
I thank A.V.M.~Herz and M.B.~Stemmler for stimulating discussions and H.~Herzel for helpful comments on the manuscript. This work was supported by the DFG through the Innovationskolleg Theoretische Biologie and the Graduiertenkolleg 120.
\end{acknowledgments}

\bibliography{/home/machens/latexfiles/references/ref}

\begin{thebibliography}{17}
\bibitem[{\citenamefont{footnote}(1)}]{footnote}
\bibinfo{author}{For a deterministic input-output system, the mutual information is given by $I=H_y$ since $H_y(x) = 0$ for all $x$. Maximizing the mutual information results in a uniform distribution of the outputs $y$. For a one-dimensional system with a monotonic relation $y=f(x)$, we have $p(x)dx = p(y)dy$ and the optimal input distribution is simply given by $p(x) \propto dy/dx$. This relation is approximately preserved in the stochastic case.}
\end{thebibliography}

\end{document}